\documentclass[a4paper]{cas-dc}

\usepackage[numbers,sort&compress]{natbib}
\usepackage{hyperref}
\hypersetup{colorlinks=false}
\usepackage{amsthm}
\usepackage{bm}
\usepackage{algorithm}
\usepackage{algorithmic}

\newtheorem{theorem}{Theorem}
\newdefinition{definition}{Definition}

\begin{document}
\let\printorcid\relax
\def\floatpagepagefraction{1}
\def\textpagefraction{.001}

\shorttitle{Identifying the Hierarchical Emotional Areas in the Human Brain Through Information Fusion}

\shortauthors{Z. Huang et al.}

\title[mode=title]{Identifying the Hierarchical Emotional Areas in the Human Brain Through Information Fusion}

\author[1,2]{Zhongyu Huang}

\ead{huangzhongyu2020@ia.ac.cn}

\credit{Conceptualization, Data curation, Formal analysis, Investigation, Methodology, Software, Visualization, Writing -- original draft}

\affiliation[1]{organization={Laboratory of Brain Atlas and Brain-Inspired Intelligence, State Key Laboratory of Multimodal Artificial Intelligence Systems, Institute of Automation, Chinese Academy of Sciences},
            city={Beijing},
            postcode={100190},
            country={China}}

\affiliation[2]{organization={School of Artificial Intelligence, University of Chinese Academy of Sciences},
            city={Beijing},
            postcode={100049},
            country={China}}

\author[1]{Changde Du}

\credit{Funding acquisition, Supervision, Validation, Writing -- review \& editing}

\author[3]{Chaozhuo Li}

\credit{Investigation, Methodology, Writing -- review \& editing}

\affiliation[3]{organization={Key Laboratory of Trustworthy Distributed Computing and Service (MoE), Beijing University of Posts and Telecommunications},
            city={Beijing},
            postcode={100876},
            country={China}}

\author[1,2]{Kaicheng Fu}

\credit{Conceptualization, Data curation}

\author[1,2]{Huiguang He}

\ead{huiguang.he@ia.ac.cn}

\credit{Funding acquisition, Project administration, Supervision, Writing -- review \& editing}

\cormark[1]

\cortext[1]{Corresponding author}

\begin{abstract}
The brain basis of emotion has consistently received widespread attention, attracting a large number of studies to explore this cutting-edge topic.
However, the methods employed in these studies typically only model the pairwise relationship between two brain regions, while neglecting the interactions and information fusion among multiple brain regions---one of the key ideas of the psychological constructionist hypothesis.
To overcome the limitations of traditional methods, this study provides an in-depth theoretical analysis of how to maximize interactions and information fusion among brain regions.
Building on the results of this analysis, we propose to identify the hierarchical emotional areas in the human brain through multi-source information fusion and graph machine learning methods.
Comprehensive experiments reveal that the identified hierarchical emotional areas, from lower to higher levels, primarily facilitate the fundamental process of emotion perception, the construction of basic psychological operations, and the coordination and integration of these operations.
Overall, our findings provide unique insights into the brain mechanisms underlying specific emotions based on the psychological constructionist hypothesis.
\end{abstract}

\begin{highlights}
\item Identify hierarchical emotional areas to study brain mechanisms underlying emotion.
\item Conduct an in-depth theoretical analysis based on information fusion and graph theory.
\item Develop a novel framework to improve emotion decoding using the identified areas.
\item Demonstrate the potential of identified areas by cross-dataset emotion decoding.
\end{highlights}


\maketitle

\section{Introduction}
\label{sec:intro}

Emotions are complex psychological and physiological states that arise in response to stimuli from internal feelings or external environments.
Generally, these external stimuli are perceived in diverse manners, such as visual experiences, audio signals, and textual information.
Our brains respond to these stimuli by generating neural activity in various brain regions, thereby producing corresponding psychological and physiological states.
Therefore, the fusion of multi-source information from either diverse manners or various brain regions contributes to understanding the brain mechanisms underlying emotion~\cite{wang2022systematic, khare2024emotion, geetha2024multimodal}.

The research on the brain basis and neural representation of emotion mainly focuses on two hypotheses: \emph{locationism} and \emph{psychological constructionism}.
The locationist hypothesis suggests that discrete emotion categories consistently and specifically correspond to distinct brain regions, implying a one-to-one mapping between brain regions and emotion categories~\cite{lindquist2012brain}.
On the contrary, the psychological constructionist hypothesis suggests that discrete emotion categories are constructed from more general brain networks, i.e., the interactions among various brain regions, rather than specific brain regions~\cite{lindquist2012brain}.
While both hypotheses are of equal age, there is relatively little evidence supporting the locationist hypothesis.
In contrast, a large number of studies~\cite{lindquist2012brain, kassam2013identifying, wager2015bayesian, saarimaki2016discrete, huang2018studying, saarimaki2018distributed, horikawa2020neural, koide2020distinct, saarimaki2022classification, xu2023functional} demonstrate that the neural representation of emotion is distributed, leading to the conclusion that the psychological constructionist hypothesis has greater potential for revealing the brain basis of emotion.

To investigate the brain basis of emotion, many previous studies have employed various methods to identify patterns of brain activity associated with specific emotions, such as multivariate pattern analyses~\cite{saarimaki2016discrete, saarimaki2018distributed}, Bayesian approaches~\cite{kassam2013identifying, wager2015bayesian}, and brain networks~\cite{huang2018studying, saarimaki2022classification, xu2023functional}.
However, these methods either fail to leverage the relationships between brain regions or only model the pairwise relationship between two brain regions, while neglecting the interactions and information fusion among multiple brain regions---one of the key ideas of the psychological constructionist hypothesis.
As a result, their findings on the brain regions involved in emotion encoding vary, and even a consistent conclusion on the brain mechanisms for specific emotions has yet to be established.

To reveal the brain mechanisms underlying emotion based on the psychological constructionist hypothesis, it is essential to consider the interactions and information fusion among multiple brain regions at a higher level.
From a network neuroscience perspective, the brain regions involved in emotion encoding present a modular structure~\cite{griffiths1990modularity, heilman1994emotion}.
Modules with similar functions are coupled together, forming a unique area with a specific function.
We refer to this unique area as the \emph{emotional area}, similar to the functional areas in the human brain.
Moreover, a series of influential studies~\cite{meunier2009hierarchical, meunier2010modular} demonstrates that brain networks further exhibit hierarchically modular organizations.
The main module in a brain network plays a dominant role, with a set of sub-modules cooperating to perform subordinate functions.
In addition to these sub-modules, there exist sub-sub-modules and so forth, each playing an increasingly subtle role.
Accordingly, in the context of emotion encoding, the entire brain network can be decomposed into hierarchical subnetworks that correspond to \emph{hierarchical emotional areas}.

Since emotions are complex subjective experiences, they include multiple psychological operations and cognitive processes, thereby involving various brain regions~\cite{lindquist2012brain}.
When an individual experiences emotional stimuli, the entire brain network is decomposed into hierarchical subnetworks.
Each of these subnetworks corresponds to an emotional area and performs distinct functions, contributing to different psychological operations and cognitive processes involved in emotion encoding.
Meanwhile, subnetworks in various emotional areas interact and collaborate to create a comprehensive brain network for emotion encoding.
Based on the hierarchically modular organization of the brain network, this analysis aligns perfectly with both the key idea of the psychological constructionist hypothesis and the professional knowledge in affective science.
Therefore, this work aims to investigate the hierarchical emotional areas in the human brain through multi-source information fusion and graph machine learning methods.
Our ultimate goal is to reveal the brain mechanisms underlying specific emotions based on the psychological constructionist hypothesis.

Overall, the key innovations and main contributions of our research can be summarized as follows:
\begin{itemize}
    \item We introduce the concept of hierarchical emotional areas to understand the brain mechanisms underlying emotion, aligning perfectly with both the key idea of the psychological constructionist hypothesis and the professional knowledge in affective science.
    \item We provide an in-depth theoretical analysis of how to maximize interactions and information fusion among brain regions. Building on the results of this analysis, we identify hierarchical emotional areas in the human brain.
    \item We develop a novel framework, Hierarchical Emotion Network (HEmoN), which improves emotion decoding tasks by exploiting the identified hierarchical emotional areas.
    \item To verify the rationality and effectiveness of our approach, we conduct experiments on two large-scale datasets, both containing audio-visual multi-source naturalistic stimuli. Furthermore, we accomplish a challenging task: identifying hierarchical emotional areas on one dataset and applying them to cross-dataset emotion decoding on another dataset.
\end{itemize}

\section{Related work}

Emotions arise from the activations of specialized neuronal populations in the human brain.
Therefore, researchers have devoted themselves to exploring the neural representation of emotion to gain a deeper understanding of emotion.
One of the most prominent research areas is \emph{emotion encoding and decoding}, which aims to identify the mapping patterns between emotional stimuli and their corresponding brain responses.
In essence, encoding and decoding are complementary operations: encoding investigates how to transform stimuli into brain responses, while decoding investigates how to predict information about the stimuli from brain responses~\cite{naselaris2011encoding}.

In laboratory settings, researchers usually measure and collect the brain signals of subjects when they experience emotional stimuli.
By analyzing the collected brain signals and the experienced emotional stimuli, researchers can delve into emotion encoding and decoding.
Commonly used brain signals include functional magnetic resonance imaging (fMRI), electroencephalography (EEG), and magnetoencephalography (MEG).
Due to its high spatial resolution, fMRI allows precise localization of the active brain regions involved in emotional processing~\cite{nguyen2019cortical}, providing an excellent opportunity to study the underlying brain mechanisms for specific emotions~\cite{gu2019integrative}.
On the other hand, traditional emotional stimuli include images, especially facial expressions~\cite{vuilleumier2001effects, sato2004enhanced}, and mental imagery~\cite{holmes2010mental, saarimaki2016discrete}.
However, these stimuli are abstract, simplified, and, in many ways, fail to reflect the complexity and dynamics inherent in real-life behaviors and stimuli~\cite{sonkusare2019naturalistic}.
To investigate real-life emotional experiences, researchers are increasingly turning to naturalistic paradigms, such as watching movies~\cite{horikawa2020neural, xu2023functional}, listening to stories~\cite{smirnov2019emotions, saarimaki2022classification}, or experiencing audio-visual multi-source stimuli~\cite{hanke2016studyforrest, lettieri2019emotionotopy, koide2020distinct}.
A growing body of evidence suggests that these \emph{naturalistic stimuli} benefit from the fusion of multi-source information, incorporate richer and more dynamic experiences, evoke stronger emotions, and provide a more reasonable approximation to freeform cognition in our daily lives~\cite{sonkusare2019naturalistic, saarimaki2021naturalistic, bernstein2022prediction, ezzameli2023emotion}.

Given their simplicity, efficiency, and strong interpretability, a large number of studies~\cite{kassam2013identifying, wager2015bayesian, saarimaki2016discrete, saarimaki2018distributed, lettieri2019emotionotopy, horikawa2020neural, koide2020distinct} employ linear regression or multivariate statistical methods to develop models for emotion encoding or decoding.
Unfortunately, these methods are unable to leverage the rich relationships between brain regions.
To overcome this limitation, some pioneering works~\cite{koelsch2014functional, huang2018studying, lin2019investigation, ghahari2020investigating, li2021braingnn, saarimaki2022classification, li2023well, liu2023decoding, xu2023functional} use fMRI data to construct functional brain networks~\cite{bullmore2009complex}, which can explicitly model the pairwise relationship between brain regions.
A series of works~\cite{koelsch2014functional, huang2018studying, lin2019investigation, ghahari2020investigating, li2023well} uses diverse network measures to assess the significance of each region of interest in the brain network, thereby studying the mechanisms of emotion encoding.
Another series of works~\cite{saarimaki2022classification, liu2023decoding, xu2023functional} maps various emotion categories into distinct brain networks with more distinguishable graph structural features, thereby improving the performance of emotion decoding.
In addition, several recent advances~\cite{huang2018studying, li2021braingnn, huang2023graph, lettieri2024dissecting} have made remarkable achievements in interpreting emotions through various graph machine learning methods.
\citet{huang2018studying} investigate the connectivity within the brain network and reveal a general emotion pathway connecting neural nodes involved in basic psychological operations.
\citet{li2021braingnn} design a graph neural network framework for analyzing brain networks and utilize it to infer salient brain regions for emotional tasks.
\citet{huang2023graph} construct an emotion-brain bipartite graph to model the relationships between emotions and brain regions, thereby improving the representation learning process.
\citet{lettieri2024dissecting} study the encoding rules of emotional instances and discover that they are represented in an extensive network encompassing sensory, prefrontal, and temporal regions.

Although these works have employed graph machine learning methods to investigate emotional information in the brain more effectively, their primary focus remains on identifying key brain regions or capturing pairwise relationships between two objects.
These approaches have essentially deviated from the key idea of the psychological constructionist hypothesis, which emphasizes the interactions and information fusion among multiple brain regions \emph{throughout} the entire methodology.
In addition, these works have rarely considered the hierarchically modular organization of the brain network, neglecting important facts in network neuroscience.
Consequently, we aim to investigate the hierarchical emotional areas in the human brain by combining graph machine learning methods with information fusion, thereby overcoming the above challenges and revealing the brain mechanisms underlying specific emotions.

\section{Preliminaries}

\subsection{Graph theory}

Let $G = (\mathcal{V}, \mathcal{E}) \in \mathcal{G}$ be a graph with vertex set $\mathcal{V} = \{v_1, v_2, \ldots, v_{N_G}\}$ and edge set $\mathcal{E} = \{e_1, e_2, \ldots, e_{M_G}\}$, where $N_G = |\mathcal{V}|$ and $M_G = |\mathcal{E}|$ represent the number of vertices and edges in $G$, respectively.
Let $\bm A \in \mathbb{R}^{N_G \times N_G}$ be the adjacency matrix of $G$, and $\bm X \in \mathbb{R}^{N_G \times c_0}$ be its node feature matrix.
For each node $ v \in \mathcal{V}$, $\bm x_v \in \mathbb{R}^{c_0}$ denotes its node feature vector, and $d_v$ denotes its degree (i.e., the number of 1-hop neighbors).
The set of $v$'s 1-hop neighbors is represented as $\mathcal{N}(v)$.

A \emph{walk} in graph $G$ is a finite sequence of alternating vertices and edges, such as $v_0, e_1, v_1, e_2, \ldots, e_m, v_m$, in which each edge $e_i = (v_{i-1}, v_i)$.
A walk may contain repeated edges.
A \emph{path} is a walk in which all vertices, and hence all edges, are distinct (except, possibly, $v_0 = v_m$).
We denote a path $P$ with length $m$ as its vertex sequence $(v_0, v_1, \ldots, v_m)$.
If $v_0 = v_m$ and $m \ge 2$, this path is called a \emph{cycle}.

In an unweighted graph $G$, the \emph{shortest path} between two vertices is the path with the minimum length between them, and its length is called the \emph{(shortest path) distance} between these two vertices.
We use $\text{dist}(u, v)$ to denote the distance between a pair of vertices $u$ and $v$.
The \emph{longest shortest path} in graph $G$ is the shortest path with the maximum length between any pair of vertices.
The \emph{diameter} of graph $G$ is the length of the longest shortest path, defined as
\begin{equation*}
    \mathfrak{d} = \max \left\{ \text{dist}(u, v) \mid \forall u, v \in \mathcal{V} \right\}.
\end{equation*}
In this paper, we use the normal $d$ to denote the node \emph{degree} and the Gothic $\mathfrak{d}$ to denote the graph \emph{diameter}.

A \emph{tree} is a special type of graph that is connected and acyclic.
Trees have several essential properties; for instance, there exists exactly one path between any pair of vertices.
Additionally, a tree with $N_T$ vertices has exactly $N_T-1$ edges.
A \emph{forest} is a collection of disjoint trees, and a tree is a special type of forest with only one connected component.

\subsection{Psychological constructionist hypothesis}

The psychological constructionist hypothesis assumes that emotion categories are common-sense categories, with instances emerging from the combination of more basic psychological operations, such as core affect, conceptualization, executive attention, language, etc.
These psychological operations are the common ingredients of all mental states.

The \emph{core affect} is a mental representation of bodily sensations, enabling an organism to determine whether something in the environment has motivational salience, e.g., whether it is beneficial, detrimental, approachable, or avoidable.
The \emph{conceptualization} is a process that imbues sensations from the body or the external world with meaning in a given context by drawing on representations of previous experiences.
These representations are activated by the current physical and psychological situation.
The \emph{executive attention} is a process that selectively enhances some representations while suppressing others.
It can regulate the activity in other processes, such as core affect, conceptualization, or language use.
The \emph{language word} serves as an “essence placeholder,” which helps integrate feelings, behaviors, and expressions into instances of a meaningful emotion category.

\section{Hierarchical emotional areas: Theoretical analysis and methodology}

\begin{figure*}
    \centering
    \includegraphics[width=0.95\textwidth]{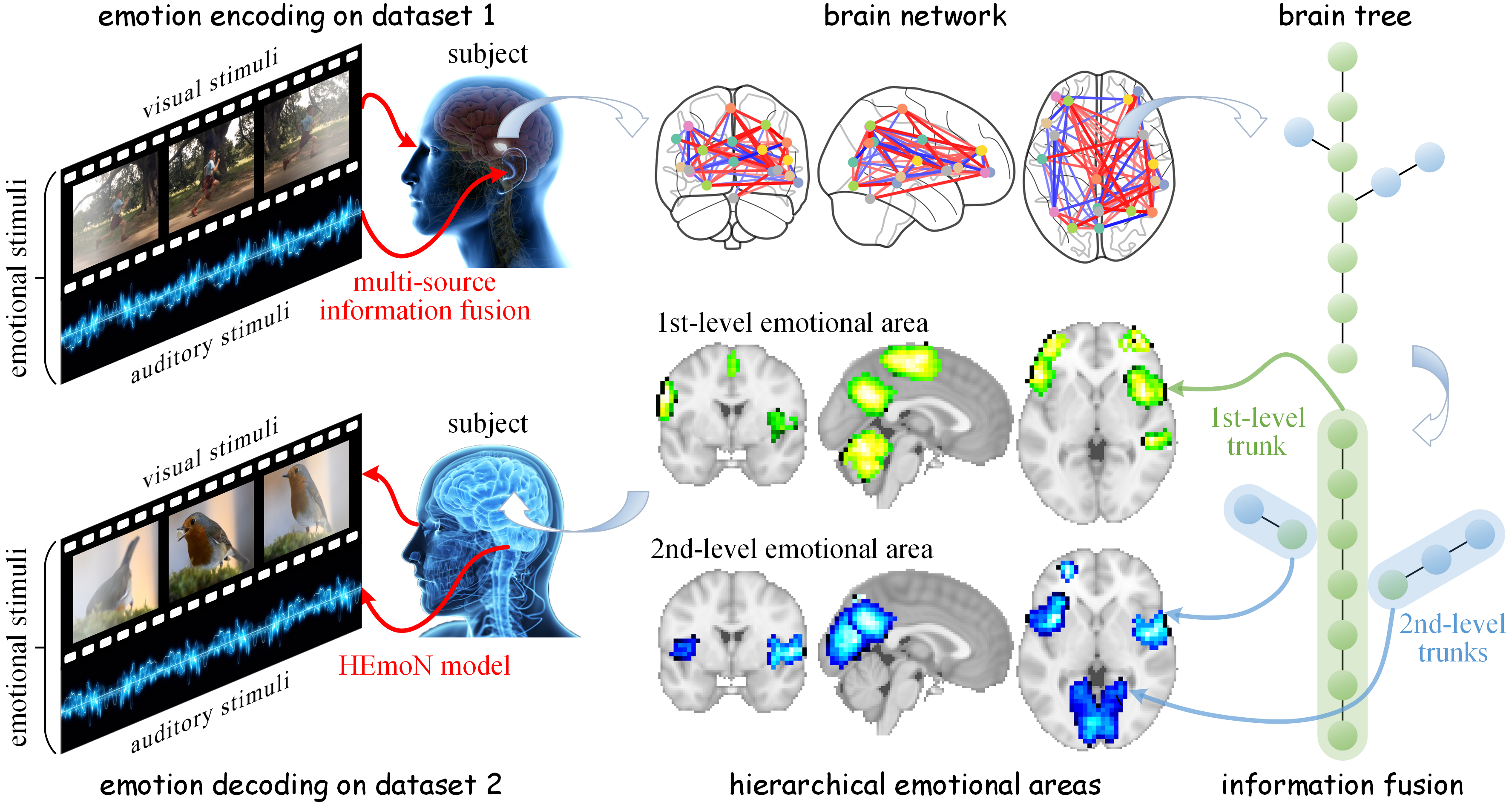}
    \caption{The pipeline of identifying hierarchical emotional areas in the human brain.
    To investigate the brain mechanisms underlying emotion (i.e., emotion encoding), we first measure and collect the brain signals of subjects when they experience audio-visual multi-source emotional stimuli.
    Next, we construct a brain network using the collected fMRI signals and extract a brain tree from this brain network.
    Then, we decompose the entire brain tree into trunks at different levels (i.e., hierarchical trunks), with each trunk facilitating information fusion among brain regions.
    Following the decomposition, we revert these hierarchical trunks in the brain tree back to hierarchical emotional areas in the human brain.
    As a result, we complete the identification of hierarchical emotional areas on a given dataset.
    Finally, we use the proposed model, HEmoN, which builds on these identified hierarchical emotional areas, to perform cross-dataset emotion decoding on other challenging datasets.}
    \label{fig:pl}
\end{figure*}

In this section, we present the identification of hierarchical emotional areas and their application in emotion decoding.
We start by extracting a brain tree from the brain network to overcome its challenges and facilitate the capture of hierarchical information.
Furthermore, we provide an in-depth theoretical analysis based on this brain tree to maximize interactions and information fusion among brain regions, aligning with the key idea of the psychological constructionist hypothesis.
Building on the results of this analysis, we identify the hierarchical emotional areas and present corresponding technical implementations.
Finally, we develop a novel framework to improve emotion decoding tasks by exploiting the identified hierarchical emotional areas.
The pipeline of this study is illustrated in Figure~\ref{fig:pl}.

\subsection{Extracting the brain tree}
\label{sec:tree}

We first construct a brain network, $G = (\mathcal{V}, \mathcal{E})$, using the collected fMRI signals.
In this network, nodes are defined as brain regions of interest (ROIs), generally derived from an existing brain parcellation, such as the Power Atlas~\cite{power2011functional} and the Destrieux Atlas~\cite{destrieux2010automatic}.
The edges are defined as the functional connectivity between these ROIs, with functional connectivity commonly computed as the pairwise correlation of the fMRI time series between two ROIs.

However, conventional brain networks may suffer from various limitations, such as threshold problems~\cite{fornito2010network, stam2014trees, tewarie2015minimum, van2018minimum}, spurious and noisy connections~\cite{fornito2010network, stam2014trees, tewarie2015minimum, van2018minimum}, scaling effects~\cite{fornito2010network, tewarie2015minimum}, and methodological biases~\cite{stam2014trees, tewarie2015minimum}.
Fortunately, many studies~\cite{stam2014trees, tewarie2015minimum, van2018minimum} have demonstrated that the spanning tree can effectively overcome these challenges and capture the essential properties of complex brain networks, leading to great success in a wide range of applications.
Moreover, previous research~\cite{sonthalia2020tree} has indicated that the tree structure is an efficient tool for capturing hierarchical information from data, aligning with our goal of identifying hierarchical emotional areas.
Therefore, we propose to extract a \emph{brain tree} from the original brain network using the spanning tree algorithm.

Specifically, given a brain network, we first sort all edges in descending order according to their weights.
The construction process of the brain tree begins by using the edge with the highest weight.
Subsequently, we select the edge with the highest weight from the remaining edges.
If the newly selected edge forms a cycle with the previously selected edges, it is discarded; otherwise, it is added to the brain tree.
This process is executed iteratively until there are $N_G-1$ edges in the brain tree.
Finally, all edge weights in the brain tree are set to a value of one.
In this way, we transform the brain network $G = (\mathcal{V}, \mathcal{E})$ into a brain tree $T = (\mathcal{V}, \mathcal{R})$, where $\mathcal{V}$ denotes the vertex set of $T$, identical to that of $G$, and $\mathcal{R}$ denotes the edge set of $T$.

\subsection{Theoretical analysis: Interactions and information fusion in the brain tree}

As discussed in Section~\ref{sec:intro}, the psychological constructionist hypothesis emphasizes the interactions and information fusion among multiple brain regions.
Since these interactions can be essentially considered as the information influence that one brain region (i.e., node) exerts on another, we start our analysis by quantitatively describing the information influence between an arbitrary pair of nodes, $u$ and $v$.
Based on previous literature~\cite{xu2018representation, huang2020graph}, we take the random walk as a general rule for information propagation in the brain tree and define node influence as follows.

\begin{definition}[Node Influence]
    Let $\bm h_v^{(0)} \in \mathbb{R}^{c_0}$ be the initial feature vector of node $v$, and $\bm h_u^{(k)}$ be the representation vector of node $u$ after $k$ steps of the random walk.
    The node influence $I(u, v)$ of node $v$ on node $u$ after $k$ steps of the random walk is defined as the norm of the Jacobian matrix $\partial \bm h_u^{(k)} / \partial \bm h_v^{(0)}$:
    \begin{equation*}
    I(u, v) = \left\lVert \frac{\partial \bm h_u^{(k)}}{\partial \bm h_v^{(0)}} \right\rVert,
	\end{equation*}
    where the norm $\Vert \cdot \Vert$ is an arbitrary induced norm\footnote{An induced norm is a matrix norm induced by a vector norm, defined as $\Vert \bm A \Vert = \sup \left\{ \Vert \bm{Ax} \Vert \mid \Vert \bm x \Vert = 1 \right\}$.}.
\end{definition}

The node influence quantifies how changes in the information of node $v$ affect that of node $u$.
The following theorem presents an analytical expression of the node influence for further analysis.

\begin{theorem}
\label{thm:influ}
    Given a tree $T$, the node influence $I_T(u, v)$ of node $v$ on node $u$ after $k = \text{dist}(u, v)$ steps of the random walk is
    \begin{equation*}
        I_T(u, v) = \left| \frac{a_{uv_1} \cdot a_{v_1v_2} \cdots a_{v_{k-1}v}}{\sum\limits_{i \in \mathcal{N}(u)} {a_{ui}} \cdot \sum\limits_{j \in \mathcal{N}(v_1)} {a_{v_1j}} \cdots \sum\limits_{l \in \mathcal{N}(v_{k-1})} {a_{v_{k-1}l}}} \right|,
	\end{equation*}
    where the only path between nodes $u$ and $v$ is denoted as $(u, v_1, v_2, \cdots, v_{k-1}, v)$;
    $a_{v_iv_j}$ is the edge weight between nodes $v_i$ and $v_j$, and $\mathcal{N}(u)$ is the set of $u$'s 1-hop neighbors.
\end{theorem}

Detailed proof of Theorem~\ref{thm:influ} is provided in Appendix~\ref{sec:pf_influ}.
Since all edges in the brain tree have a uniform weight of~1, the result of Theorem~\ref{thm:influ} can be further simplified to
\begin{equation}
\label{eq:IT}
    I_T(u, v) = \frac{1}{d_u d_{v_1} \cdots d_{v_{k-1}}}
\end{equation}
where $d_u$ is the degree of node $u$.
Eq.~\eqref{eq:IT} indicates that node influence decreases as the distance $k$ between nodes and node degrees increase, consistent with intuition.
More precisely, node influence $I_T(u, v)$ decreases \emph{exponentially} with the increase in the distance between nodes $u$ and $v$, and decreases \emph{polynomially} with the increase in node degrees on the path connecting $u$ and $v$.

To facilitate effective interactions among brain regions, we aim to maximize the node influence $I_T(u, v)$ in the brain tree.
This objective can be achieved by solving the following optimization problem:
\begin{align*}
    \max_{d_u, d_{v_i}} \quad &
    I_T(u, v) = \frac{1}{d_u d_{v_1} \cdots d_{v_{k-1}}} \\
    \text{s.t.} \quad &
    d_u \phantom{\;} \ge 1 \\
    &
    d_{v_i} \ge 2, \enspace \forall i = 1, 2, \ldots, k-1
\end{align*}
Clearly, the optimal solution is $\hat{d}_u = 1, \hat{d}_{v_i} = 2$, which corresponds to the path between nodes $u$ and $v$.
Since trees have an essential property that there exists \emph{exactly} one path between any pair of nodes, the path between $u$ and $v$ is also the shortest path.
Accordingly, the optimal solution motivates us to propagate information along the shortest path, thereby facilitating interactions among brain regions.

Furthermore, it is essential to consider the fusion of information among multiple brain regions, enabling them to form a unique emotional area with a specific function.
The above analysis indicates that information should be propagated along the (shortest) path.
Consequently, we continue our analysis by quantifying the total amount of information carried by a path and defining path information as follows.

\begin{definition}[Path Information]
    Let $P$ denote a path of length $m$ with vertex sequence $(v_0, v_1, \ldots, v_m)$.
    The path information $I(P)$ of path $P$ is defined as the sum of the node influences between every pair of nodes on the path:
    \begin{equation*}
    I(P) = \sum_{\substack{i,j=0 \\ i \neq j}}^m {I_P(v_i, v_j)}.
	\end{equation*}
\end{definition}

The path information quantifies the extent of information fusion among nodes on path $P$.
We aim to maximize the path information, thereby enhancing the fusion of information among brain regions as much as possible.
The following theorem gives the path with the maximum path information.

\begin{theorem}
\label{thm:diam}
    Given a tree $T$, the longest shortest path $\hat{P}$ has the maximum path information
    \begin{equation*}
    I(\hat{P}) = \sum_{k=1}^\mathfrak{d} {(\mathfrak{d}-k+1) \cdot \frac{2}{2^{k-1}}},
	\end{equation*}
    where $\mathfrak{d}$ is the diameter of tree $T$.
\end{theorem}

Detailed proof of Theorem~\ref{thm:diam} is provided in Appendix~\ref{sec:pf_diam}.
Theorem~\ref{thm:diam} indicates that the longest shortest path has the maximum amount of information among all possible paths in a brain tree.
Consequently, the theoretical analysis in this subsection demonstrates that propagating information along the longest shortest path on a brain tree maximizes interactions and information fusion among brain regions.
In the next subsection, we will exploit the longest shortest path to further identify the emotional area and present the corresponding implementations.

\subsection{Identifying the hierarchical emotional areas}
\label{sec:area}

In the natural world, the tree trunk is an essential component of a tree.
The remaining parts of a tree after removing the tree trunk are also regarded as (sub)trees, each with its own “(sub)tree trunk.”
Thus, natural tree trunks exhibit hierarchical organization throughout the entire tree.
Inspired by this interesting natural phenomenon, we decompose the entire brain tree into hierarchical trunks and leverage them to identify the hierarchical emotional areas.

Specifically, the 1st-level trunk, $\mathfrak{t}_1$, serves as the main trunk of the entire brain tree, and there is usually only one main trunk at the first level.
The trunk at the remaining $\ell$th ($\ell \ge 2$) level, $\mathfrak{t}_\ell$, is considered a branch of the entire brain tree, and there are usually multiple branches at the $\ell$th level.
We denote the set of trunks at the $\ell$th level as $\mathcal{T}_\ell$.
Our previous theoretical analysis suggests that propagating information along the longest shortest path on a brain tree maximizes interactions and information fusion among brain regions.
Therefore, we designate the longest shortest path as the main trunk $\mathfrak{t}_1$.
In addition, the brain regions corresponding to all nodes on this main trunk constitute the main emotional area, denoted as $\mathcal{A}_1$.
Once the main trunk is obtained and removed from the brain tree, the remaining part becomes a forest.
This forest contains at least one (often multiple) connected component(s), and every component is a (sub)tree~\cite{west2001introduction}.
To identify the trunks at the $\ell$th ($\ell \ge 2$) level, we search for the longest shortest path in each connected component of the forest after removing the trunks of all previous $(\ell-1)$ levels.
Similarly, the brain regions corresponding to all nodes on these $\ell$th-level trunks constitute the $\ell$th-level emotional area, denoted as $\mathcal{A}_\ell$.
This process runs \emph{iteratively} and ends once all nodes in the tree have been removed.
Figure~\ref{fig:example} in Appendix~\ref{sec:example} provides a concrete example to illustrate the process above.

\begin{algorithm}
\caption{Identifying the hierarchical emotional areas.}
\label{algo:area}
\begin{algorithmic}[1]
    \renewcommand{\algorithmicrequire}{\textbf{Input:}}
    \renewcommand{\algorithmicensure}{\textbf{Output:}}
    \REQUIRE A brain tree $T = (\mathcal{V}, \mathcal{R})$.
    \ENSURE The emotional areas at every level.
    \STATE $\ell \gets 0$;
    \STATE $F_0 \gets (\mathcal{V}, \mathcal{R})$;
    \WHILE{$\mathcal{V} \ne \emptyset$}
        \STATE $\ell \gets \ell+1$;
        \FOR{$\mathfrak{c} = 1, \ldots, \mathfrak{C}_{\ell-1}$}
            \STATE Identify an $\ell$th-level trunk $\mathfrak{t}_{\ell,\mathfrak{c}}$ from the $\mathfrak{c}$th connected component of $F_{\ell-1}$;
            \STATE Get the set $\mathcal{A}_{\ell,\mathfrak{c}}$ of all nodes in $\mathfrak{t}_{\ell,\mathfrak{c}}$;
            \STATE Get the set $\mathcal{R}_{\ell,\mathfrak{c}}$ of all edges in $\mathfrak{t}_{\ell,\mathfrak{c}}$;
            \STATE $\mathcal{R} \gets \mathcal{R} \setminus \mathcal{R}_{\ell,\mathfrak{c}}$;
            \STATE Find the set $\mathcal{V}_{\ell,\mathfrak{c}}$ of all isolated nodes;
            \STATE $\mathcal{V} \gets \mathcal{V} \setminus \mathcal{V}_{\ell,\mathfrak{c}}$;
        \ENDFOR
        \STATE $\mathcal{T}_\ell \gets \{\mathfrak{t}_{\ell,1}, \ldots, \mathfrak{t}_{\ell,\mathfrak{C}_{\ell-1}}\}$;
        \STATE $\mathcal{A}_\ell \gets \mathcal{A}_{\ell,1} \cup \cdots \cup \mathcal{A}_{\ell,\mathfrak{C}_{\ell-1}}$;
        \STATE $F_\ell \gets (\mathcal{V}, \mathcal{R})$;
    \ENDWHILE
    \RETURN $\mathcal{A}_1, \mathcal{A}_2, \cdots, \mathcal{A}_L$
\end{algorithmic}
\end{algorithm}

Suppose there are $L$ levels in total.
Algorithm~\ref{algo:area} describes the detailed process of identifying hierarchical emotional areas.
Here, $F_\ell$ represents a forest obtained after removing the trunks of all previous $\ell$ levels and the isolated nodes.
Line~2 initializes the forest $F_0$ to tree $T = (\mathcal{V}, \mathcal{R})$.
Line~5 considers all $\mathfrak{C}_{\ell-1}$ connected component(s) in $F_{\ell-1}$.
Line~6 searches for the longest shortest path in the $\mathfrak{c}$th connected component of $F_{\ell-1}$ to identify an $\ell$th-level trunk $\mathfrak{t}_{\ell,\mathfrak{c}}$.
Lines~7 and 14 identify the emotional area, $\mathcal{A}_\ell$, at the $\ell$th level.
Lines~8-9 remove all edges in $\mathfrak{t}_{\ell,\mathfrak{c}}$ from $F_{\ell-1}$, and Lines~10-11 remove all isolated nodes.
Line~15 updates $F_\ell$ based on the results of Lines~9 and 11.
If there are no remaining nodes, Line~16 terminates the loop, and finally, Line~17 returns the emotional areas at every level.

\subsection{Practical application: Emotion decoding}
\label{sec:app}

We further develop a novel framework, Hierarchical Emotion Network (HEmoN), to improve emotion decoding tasks by exploiting the identified hierarchical emotional areas.
As previously analyzed, we identify hierarchical trunks by iteratively searching for the longest shortest path.
Given that nodes on the longest shortest path (i.e., a trunk) form a long sequence, we use the Long Short-Term Memory (LSTM)~\cite{hochreiter1997long} to effectively capture long-term dependencies and recognize complex patterns, thereby facilitating the representation learning process of emotion decoding.
Specifically, after obtaining the trunks at every level according to Line~13 of Algorithm~\ref{algo:area}, we apply an LSTM to learn the representation for each trunk along its corresponding trunk path, formulated as
\begin{equation}
\label{eq:trunk}
    \bm h_T^{(\ell)} = \sum_{\mathfrak{t}_\ell \in \mathcal{T}_\ell} {\text{LSTM} \left( \bm x_{v_0}^{(\mathfrak{t}_\ell)}, \bm x_{v_1}^{(\mathfrak{t}_\ell)}, \cdots, \bm x_{v_{k_{\mathfrak{t}_\ell}}}^{(\mathfrak{t}_\ell)} \right)}
\end{equation}
where $\mathcal{T}_\ell$ is the set of trunks at the $\ell$th level,
and $\mathfrak{t}_\ell$ is a single trunk in $\mathcal{T}_\ell$;
let the length of $\mathfrak{t}_\ell$ be $k_{\mathfrak{t}_\ell}$;
$\bm x_{v_i}^{(\mathfrak{t}_\ell)} \in \mathbb{R}^{c_0}$ denotes the feature vector of node $v_i$ on trunk $\mathfrak{t}_\ell$,
and $\bm h_T^{(\ell)} \in \mathbb{R}^{c_\ell}$ denotes the trunk representation at the $\ell$th level.
Since nodes on all trunks in $\mathcal{T}_\ell$ constitute the $\ell$th-level emotional area, $\bm h_T^{(\ell)}$ also serves as the representation of the emotional area at the $\ell$th level.
Then, we combine all the trunk representations to create a representation of the brain tree:
\begin{equation}
\label{eq:final}
    \bm h_T = \sum_{\ell=1}^L {\bm W^{(\ell)} \bm h_T^{(\ell)}}
\end{equation}
where $\bm W^{(\ell)}$ represents a learnable weighting matrix for the $\ell$th level, optimized along with other model weights during the model training process.
For emotion classification tasks, the brain tree is associated with an emotion category label $y_T$ from $C$ emotion categories.
Accordingly, we set $\bm W^{(\ell)} \in \mathbb{R}^{C \times c_\ell}$ and further input $\bm h_T$ into the softmax function to obtain the predicted class.
For emotion regression tasks, the brain tree is associated with an emotion rating vector $\bm y_T \in [0,a]^C$ containing $C$ emotion categories, where $a$ represents the maximum rating and is usually set to 100.
Accordingly, we set $\bm W^{(\ell)} \in \mathbb{R}^{C \times c_\ell}$ and further input $\bm h_T$ into the logistic function to obtain the predicted ratings.
The above representation learning process is illustrated in Appendix~\ref{sec:example} through a concrete example.

\section{Experiments}

In this section, we conduct comprehensive experiments on two large-scale datasets, both containing audio-visual multi-source naturalistic stimuli.
The entire experimental evaluation consists of two stages.
In the first stage (Section~\ref{sec:id}), we identify hierarchical emotional areas based on the overall experience of all emotions in Dataset~1~\cite{lettieri2019emotionotopy}.
In the second stage (Section~\ref{sec:decoding}), we use the proposed model, HEmoN, which builds on the identified hierarchical emotional areas, to perform cross-dataset emotion decoding on Dataset~2~\cite{koide2020distinct}.
Furthermore, we provide a thorough analysis and discussion of the experimental results in Section~\ref{sec:disc}.
Our code is publicly available at \url{https://github.com/zhongyu1998/HEmoN}.

\begin{figure*}
    \centering
    \includegraphics[width=0.955\textwidth]{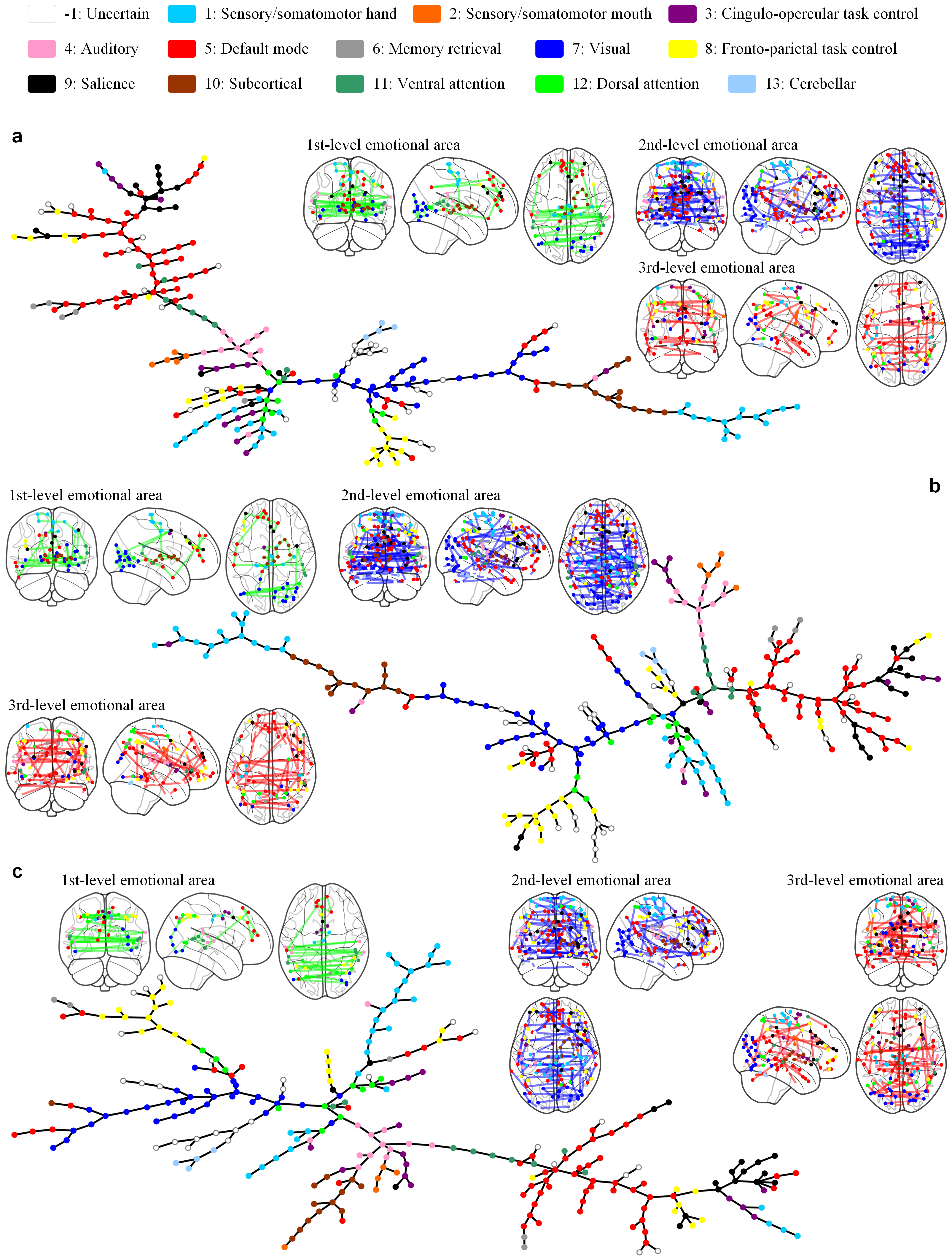}
    \caption{The illustration of the identified hierarchical emotional areas, including (a)~all basic emotions, (b)~happiness, and (c)~sadness.
    The legend at the top lists all 13 functional systems and one uncertain system, as proposed by the Power Atlas~\cite{power2011functional}.
    Each subfigure shows the corresponding brain tree and presents the emotional areas at the first three levels.
    The nodes are colored according to the legend, and the internal connections within the 1st-level, 2nd-level, and 3rd-level emotional areas are highlighted in green, blue, and red, respectively.}
    \label{fig:area}
\end{figure*}

\subsection{Dataset description}

We conduct comprehensive experiments on two large-scale datasets~\cite{lettieri2019emotionotopy, koide2020distinct} to demonstrate the effectiveness and superiority of our proposed approach.
Both datasets contain audio-visual multi-source naturalistic stimuli and use a similar experimental paradigm for data acquisition.
Specifically, they collect fMRI responses when subjects watch emotionally charged movies (i.e., emotional stimuli).
To obtain the corresponding emotion ratings for these stimuli, they recruit multiple human raters who are not involved in the fMRI experiments.
These raters label each emotional stimulus according to specific emotion categories, assigning a score to each emotion category for every stimulus.
The final emotion rating for a specific emotion category of a stimulus is determined by averaging the scores across these raters.
As for fMRI responses, we use the fMRI data preprocessed by \citet{lettieri2019emotionotopy} and \citet{koide2020distinct}.
Here, we provide an overview of the basic information and fMRI data preprocessing steps for each dataset.
Further details on emotion ratings, fMRI data acquisition, and fMRI data preprocessing are available in their original literature.

\paragraph{Dataset~1}
This dataset contains fMRI responses collected from 15 subjects when they experience emotional stimuli, featuring 3599 pairs of \{fMRI responses, emotion ratings\} for each subject.
The emotional stimuli are composed of movie segments from Forrest Gump, sourced from the StudyForrest project (\url{http://studyforrest.org}), and contain six basic emotions~\cite{ekman1992argument}: happiness, surprise, fear, sadness, anger, and disgust.
The fMRI data are preprocessed using the Advanced Normalization Tools (ANTs)~\cite{avants2009advanced} and the Analysis of Functional NeuroImages (AFNI)~\cite{cox1996afni}.
This process mainly involves the following steps:
spike correction, slice timing correction, motion correction, rigid body transformation, registration to T1-weighted images, phase distortion correction, spatial smoothing with a Gaussian kernel, and physiological noise filtering.

\paragraph{Dataset~2}
This dataset contains fMRI responses collected from 8 subjects when they experience emotional stimuli, featuring 5400 pairs of \{fMRI responses, emotion ratings\} for each subject.
The emotional stimuli are composed of movie clips sourced from the video-sharing site Vimeo (\url{https://vimeo.com/jp}) and contain 80 emotion categories.
The complete list of these 80 emotion categories and their sources are available in the original literature~\cite{koide2020distinct}.
The fMRI data are preprocessed using the Statistical Parameter Mapping toolbox (SPM8, \url{https://www.fil.ion.ucl.ac.uk/spm/software/spm8/}).
This process mainly involves the following steps:
motion correction, z-score normalization, median filter convolution, long-term trend removal, and registration to T1-weighted images.

\subsection{Implementation details}

We initialize the learnable weighting matrix in Eq.~\eqref{eq:final} and all model weights using the Xavier initializer~\cite{glorot2010understanding}.
Then, we pre-map the emotional stimuli into 64 dimensions with a linear transformation and use a batch of 32 emotional stimuli as the model input.
Our HEmoN model consists of 3 LSTM layers (Eq.~\eqref{eq:trunk}) and a final combination layer (Eq.~\eqref{eq:final}).
Each LSTM layer contains 64 hidden units, with a dropout~\cite{srivastava2014dropout} ratio of 0.2 applied to the outputs of each LSTM layer except the last one.
The model is trained using the Adam optimizer~\cite{kingma2015adam} with an initial learning rate of 0.0005.
If the performance on the validation set does not improve after 10 epochs, the learning rate is reduced by a factor of 0.5.
The training is stopped when the learning rate reaches the minimum value of 2E-5 or when the number of epochs reaches the maximum value of 300.

Our model is implemented with Python version 3.6, PyTorch version 1.4.0~\cite{paszke2019pytorch}, and PyTorch Geometric (PyG) version 1.7.0~\cite{fey2019fast}.
All experiments are conducted on a Linux server equipped with an Intel\textsuperscript{\textregistered} Xeon\textsuperscript{\textregistered} CPU E5-2650 v4 and 8 NVIDIA TITAN RTX GPUs.

\subsection{Identification of hierarchical emotional areas}
\label{sec:id}

To identify the hierarchical emotional areas, we conduct experiments on Dataset~1 and use the Power Atlas~\cite{power2011functional} to define ROIs.
The Power Atlas comprises 264 putative functional areas (i.e., ROIs) distributed across 13 functional systems and one uncertain system.
All the systems are listed at the top of Figure~\ref{fig:area}.
The fMRI time series for each ROI is obtained by averaging the activity of voxels within a fixed-radius sphere of 5mm, centered on the ROI coordinates defined by the Power Atlas.
Subsequently, we construct the brain network, extract the brain tree, and identify hierarchical emotional areas according to the methodologies described in Sections~\ref{sec:tree} and \ref{sec:area}.
We present the identification results for the overall experience of all basic emotions, as well as those for the single experiences of the most representative positive and negative emotions: happiness and sadness.
The identified hierarchical emotional areas for these three types are illustrated in Figure~\ref{fig:area}.

\subsection{Performance of emotion decoding}
\label{sec:decoding}

To perform cross-dataset emotion decoding, we build the HEmoN model by exploiting the identified hierarchical emotional areas and evaluate its performance on Dataset~2.
Let $\bm y_i$ and $\hat{\bm y}_i$ denote the true and predicted emotion rating vectors for the $i$th stimulus, respectively, where $\hat{\bm y}_i$ is calculated according to the steps described in Section~\ref{sec:app}.
Additionally, let the scalars $y_{ij}$ and $\hat{y}_{ij}$ denote the true and predicted ratings for the $j$th emotion category in the $i$th stimulus, respectively.
We take the Mean Absolute Error (MAE), which is widely used in regression tasks, as our evaluation metric.
The MAE value $\alpha$ is calculated as:
\begin{equation}
\label{eq:MAE}
    \alpha = \frac{1}{N_s} \sum_{i=1}^{N_s} {\sum_{j=1}^C {\left| y_{ij} - \hat{y}_{ij} \right|}}
\end{equation}
where $N_s$ denotes the number of stimuli.
Since most emotion categories have ratings close to or containing zeros, Eq.~\eqref{eq:MAE} performs the sum operation $\sum_{j=1}^C$ instead of the average operation $\frac{1}{C} \sum_{j=1}^C$ across all $C$ emotion categories.

\begin{figure*}
    \centering
    \includegraphics[width=0.98\textwidth]{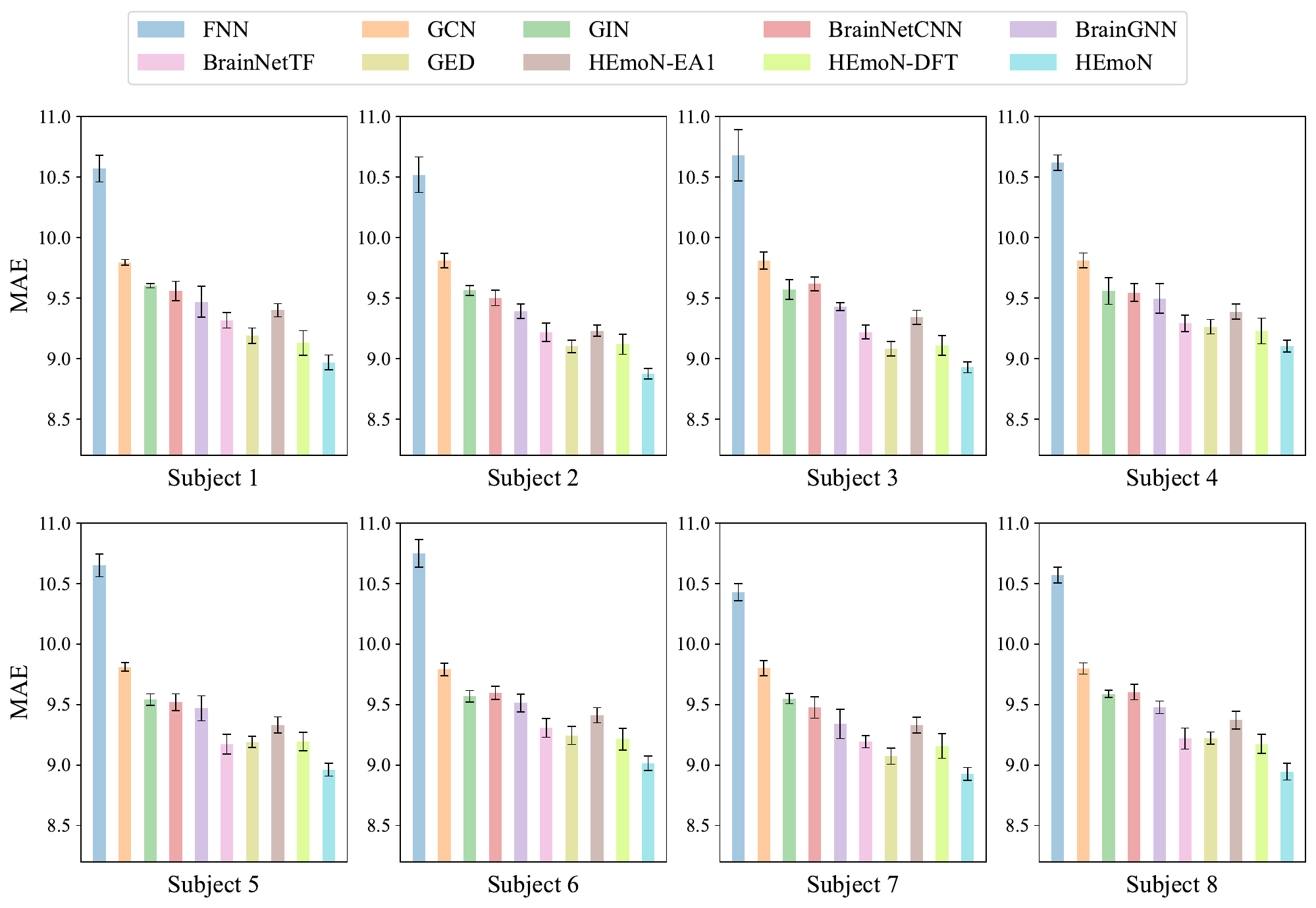}
    \caption{Emotion decoding results (measured by MAE, lower is better) on Dataset~2. The black error bars represent the standard deviation.}
    \label{fig:MAE}
\end{figure*}

Since Dataset~2 has already been split into training and test sets, we adopt the default split and perform emotion decoding between the fMRI responses to each emotional stimulus and its corresponding emotion ratings.
Specifically, in each experimental run, we use the training set consisting of 3600 pairs to train the model and then evaluate its performance on the test set consisting of 1800 pairs.
We repeat this process 10 times and report the evaluation result based on the test MAEs of these 10 runs.
Furthermore, we compare HEmoN with multiple competitive baselines, including
Fully-connected Neural Network (FNN),
Graph Convolutional Network (GCN)~\cite{kipf2017semi},
Graph Isomorphism Network (GIN)~\cite{xu2019powerful},
BrainNetCNN~\cite{kawahara2017brainnetcnn},
BrainGNN~\cite{li2021braingnn},
Brain Network Transformer (BrainNetTF)~\cite{kan2022brain},
and Graph-enhanced Emotion Decoding (GED)~\cite{huang2023graph}.
The evaluation results on Dataset~2 are summarized in Figure~\ref{fig:MAE}.

\subsection{Ablation study}

The hierarchical emotional areas and the information fusion rule of brain regions play pivotal roles in our HEmoN model.
Therefore, we conduct ablation studies to investigate their impacts on model performance.
Specifically, we replace the hierarchical emotional areas with a single \emph{1st-level emotional area} to assess the impact of hierarchical emotional areas.
In addition, we perform a \emph{depth-first traversal} on the brain tree and fuse information among brain regions based on the traversal result instead of hierarchical emotional areas to assess the impact of the information fusion rule.
The corresponding results are shown in Figure~\ref{fig:MAE}, where the suffixes “-EA1” and “-DFT” represent the 1st-level emotional area and the depth-first traversal, respectively.

\section{Discussion}
\label{sec:disc}

Figure~\ref{fig:area}a reveals that for the overall experience of all basic emotions, the 1st-level emotional area mainly includes sensory/somatomotor, auditory, default mode, visual, and subcortical systems.
These functional systems involve various senses, including sight (visual), sound (auditory), and touch (sensory/somatomotor), enabling individuals to perceive external emotional stimuli and thus creating necessary conditions for the construction of core affect.
In addition, a small portion of this area includes salience and ventral attention systems, which also contribute to the perception of external stimuli and the construction of core affect.
Moreover, the default mode and subcortical systems play key roles in distinguishing experiences of discrete emotions~\cite{wager2015bayesian, saarimaki2016discrete, barrett2017theory, horikawa2020neural, saarimaki2022classification}, thereby promoting the recognition, identification, and perception of emotions.
In summary, the 1st-level emotional area primarily facilitates the fundamental process of emotion perception.
The 2nd-level emotional area mainly includes sensory/somatomotor, cingulo-opercular task control, auditory, default mode, visual, fronto-parietal task control, salience, subcortical, and cerebellar systems.
This emotional area primarily facilitates the construction of basic psychological operations.
For instance, the anterior cingulate cortex in the cingulo-opercular task control system and the amygdala in the subcortical system both play pivotal roles in core affect~\cite{barrett2009affect, lindquist2012brain, wilson2013neural}.
In addition, a small portion of this area includes memory retrieval, ventral attention, and dorsal attention systems, which contribute to the construction of conceptualization and executive attention.
Moreover, Broca’s and Wernicke’s areas contained in this area are essential for the production and comprehension of human language~\cite{aboitiz1997evolutionary, fadiga2009broca, ardila2016localized}.
It is worth noting that our finding of the 2nd-level emotional area is consistent with a previous study~\cite{huang2018studying}, which identified a general emotion pathway connecting neural nodes involved in these basic psychological operations.
The 3rd-level emotional area mainly includes sensory/somatomotor, cingulo-opercular task control, default mode, visual, fronto-parietal task control, salience, and dorsal attention systems.
These functional systems typically participate in advanced cognitive processes, thereby facilitating the coordination and integration of the above basic psychological operations.

Figure~\ref{fig:area}b reveals that for a single experience of happiness, the 1st-level emotional area mainly includes sensory/somatomotor, default mode, visual, and subcortical systems.
In addition, a small portion of this area includes salience and ventral attention systems.
These functional systems are similar to those found in all basic emotions, with the exception of the auditory system.
A possible explanation for the absence of the auditory system is that visual information is more dominant than auditory information in the perception of happiness.
This conjecture is supported by an interesting study~\cite{most2009auditory}, which demonstrates that happiness is more easily perceived visually than auditorily.
The 2nd-level emotional area mainly includes sensory/somatomotor, cingulo-opercular task control, default mode, visual, fronto-parietal task control, salience, subcortical, ventral attention, and cerebellar systems.
In addition, a small portion of this area includes auditory, memory retrieval, and dorsal attention systems.
The activation of happiness in the brain is associated with multiple brain regions within these functional systems, such as the anterior cingulate cortex, prefrontal cortex, and insula in cingulo-opercular task control and salience systems~\cite{kringelbach2009towards, suardi2016neural}.
The 3rd-level emotional area mainly includes sensory/somatomotor, auditory, default mode, visual, fronto-parietal task control, salience, ventral attention, and dorsal attention systems.

Figure~\ref{fig:area}c reveals that for a single experience of sadness, the 1st-level emotional area mainly includes auditory, default mode, memory retrieval, visual, fronto-parietal task control, ventral attention, and dorsal attention systems.
These functional systems differ significantly from those found in all basic emotions.
Compared with the basic emotions, these functional systems incorporate memory retrieval, fronto-parietal task control, ventral attention, and dorsal attention systems, while excluding sensory/somatomotor and subcortical systems.
This distinction may stem from the fact that sadness can impair our attention, memory, and decision-making abilities~\cite{everaert2014attention, hubbard2016depressive}.
As a result, the above functional systems are the first to be affected when perceiving sadness.
The 2nd-level emotional area mainly includes sensory/somatomotor, cingulo-opercular task control, auditory, default mode, visual, fronto-parietal task control, salience, and subcortical systems.
The activation of sadness in the brain is associated with multiple brain regions within these functional systems, such as the anterior cingulate cortex, amygdala, and hippocampus in cingulo-opercular task control and subcortical systems~\cite{yoshino2010sadness, arias2020neuroscience}.
The 3rd-level emotional area mainly includes sensory/somatomotor, cingulo-opercular task control, auditory, default mode, visual, fronto-parietal task control, salience, subcortical, ventral attention, dorsal attention, and cerebellar systems.

Furthermore, the emotion decoding results in Figure~\ref{fig:MAE} show that our proposed model, HEmoN, consistently outperforms advanced models such as BrainNetTF and GED, achieving outstanding performance in all cases.
Competitive baselines, including BrainNetCNN, BrainGNN, and BrainNetTF, utilize convolutional neural networks, graph neural networks, and transformers to analyze brain networks, respectively.
However, they all neglect the information fusion among multiple brain regions.
Therefore, the superior performance of HEmoN in emotion decoding tasks highlights the importance of integrating information from multiple brain regions.
In addition, as expected, we observe an obvious performance degradation in each ablation of HEmoN-EA1 and HEmoN-DFT, demonstrating the effectiveness of both the hierarchical emotional areas and the proposed information fusion rule.
Interestingly, HEmoN-EA1 is inferior to HEmoN-DFT in all cases, further showing the potential of the identified hierarchical emotional areas.

Overall, the experimental evaluation of cross-dataset emotion decoding demonstrates the significant potential and strong generalization capability of the identified hierarchical emotional areas.
However, it is worth noting the individual differences in decoding performance among various subjects.
Investigating and addressing these individual differences constitutes an important topic for future research.

\section{Conclusion}

In this paper, we investigate the hierarchical emotional areas in the human brain through multi-source information fusion and graph machine learning methods.
Furthermore, we provide preliminary insights into the brain mechanisms underlying specific emotions based on the psychological constructionist hypothesis.
To gain a deeper understanding of the roles of emotional areas at various levels in emotion encoding, it is promising to conduct meta-analyses using a large number of existing literature in future research.

\printcredits

\section*{Declaration of competing interest}

The authors declare that they have no known competing financial interests or personal relationships that could have appeared to influence the work reported in this paper.

\section*{Data availability}

The data are publicly available on the Open Science Framework (OSF) at \url{https://osf.io/tzpdf}.

\section*{Acknowledgments}

The authors would like to express their sincere gratitude to Prof.~Dan Zhang for his helpful discussions and valuable suggestions, which greatly contributed to improving the overall quality of this paper.
They also extend their thanks to Giada Lettieri, Naoko Koide-Majima, and Shinji Nishimoto for providing the essential data for this research.

\appendix
\numberwithin{equation}{section}

\section{Proof of Theorem~\ref{thm:influ}}
\label{sec:pf_influ}

\begin{proof}
According to the propagation rule, we have
\begin{equation}
\label{eq:hu}
    \bm h_u^{(k)} = \sum_{w \in \mathcal{N}(u)} {\left( \frac{a_{uw}}{\sum_{i \in \mathcal{N}(u)} {a_{ui}}} \bm h_w^{(k-1)} \right)}
\end{equation}
where $a_{uw}$ is the edge weight between nodes $u$ and $w$, and $\mathcal{N}(u)$ is the set of $u$'s 1-hop neighbors.

After $k$ iterations of expansion as in Eq.~\eqref{eq:hu}, $\bm h_u^{(k)}$ can be represented as
\begin{align}
\label{eq:expand}
\bm h_u^{(k)}
    &= \frac{1}{\sum_{i \in \mathcal{N}(u)} {a_{ui}}} \sum_{w \in \mathcal{N}(u)} {a_{uw} \cdot \frac{1}{\sum_{j \in \mathcal{N}(w)} {a_{wj}}} \sum_{x \in \mathcal{N}(w)} {a_{wx} \cdots}} \nonumber \\
    &\phantom{{}={}} \frac{1}{\sum_{l \in \mathcal{N}(y)} {a_{yl}}} \sum_{z \in \mathcal{N}(y)} {a_{yz} \bm h_z^{(0)}}
\end{align}

Abbreviate the node influence $I_T(u, v)$ as $I$.
By substituting Eq.~\eqref{eq:expand} into the definition of node influence, we obtain
\begin{align}
I
    &= \left\lVert \frac{\partial \bm h_u^{(k)}}{\partial \bm h_v^{(0)}} \right\rVert \nonumber \\
    &= \left\lVert \frac{\partial}{\partial \bm h_v^{(0)}} \Bigg( \frac{1}{\sum_{i \in \mathcal{N}(u)} {a_{ui}}} \sum_{w \in \mathcal{N}(u)} {a_{uw} \cdot \frac{1}{\sum_{j \in \mathcal{N}(w)} {a_{wj}}} \sum_{x \in \mathcal{N}(w)} {a_{wx} \cdots}} \right. \nonumber \\
    & \phantom{= \left\lVert \frac{\partial}{\partial \bm h_v^{(0)}} \bigg( \right.}
    \left. \frac{1}{\sum_{l \in \mathcal{N}(y)} {a_{yl}}} \sum_{z \in \mathcal{N}(y)} {a_{yz} \bm h_z^{(0)}} \Bigg) \right\rVert \nonumber \\
    \label{eq:elim}
    &= \left\lVert \frac{\partial}{\partial \bm h_v^{(0)}} \left( \frac{1}{\sum_{i \in \mathcal{N}(u)} {a_{ui}}} a_{uv_1} \cdot \frac{1}{\sum_{j \in \mathcal{N}(v_1)} {a_{v_1j}}} a_{v_1v_2} \cdots \right.\right. \nonumber \\
    & \phantom{= \left\lVert \frac{\partial}{\partial \bm h_v^{(0)}} \bigg( \right.}
    \left.\left. \frac{1}{\sum_{l \in \mathcal{N}(v_{k-1})} {a_{v_{k-1}l}}} a_{v_{k-1}v} \bm h_v^{(0)} \right) \right\rVert \\
    &= \left\lVert \frac{\partial \bm h_v^{(0)}}{\partial \bm h_v^{(0)}} \right\rVert \cdot \left| \frac{a_{uv_1}}{\sum\limits_{i \in \mathcal{N}(u)} {a_{ui}}} \cdot \frac{a_{v_1v_2}}{\sum\limits_{j \in \mathcal{N}(v_1)} {a_{v_1j}}} \cdots \frac{a_{v_{k-1}v}}{\sum\limits_{l \in \mathcal{N}(v_{k-1})} {a_{v_{k-1}l}}} \right| \nonumber \\
    \label{eq:influ}
    &= \left| \frac{a_{uv_1} \cdot a_{v_1v_2} \cdots a_{v_{k-1}v}}{\sum\limits_{i \in \mathcal{N}(u)} {a_{ui}} \cdot \sum\limits_{j \in \mathcal{N}(v_1)} {a_{v_1j}} \cdots \sum\limits_{l \in \mathcal{N}(v_{k-1})} {a_{v_{k-1}l}}} \right|
\end{align}

According to the properties of trees, there exists exactly one path between nodes $u$ and $v$; we denote this path as $P = (u, v_1, v_2, \cdots, v_{k-1}, v)$.
Eq.~\eqref{eq:elim} holds because the partial derivative of nodes that are not in $P$ becomes 0, leading to their elimination.
Eq.~\eqref{eq:influ} holds because the Jacobian matrix $\partial \bm h_v^{(0)} / \partial \bm h_v^{(0)}$ is the identity matrix $I_{c_0}$, and its induced norm $\Vert \bm I_{c_0} \Vert$ equals 1.
\end{proof}

\section{Proof of Theorem~\ref{thm:diam}}
\label{sec:pf_diam}

\begin{proof}
Without loss of generality, we consider an arbitrary path $P$ of length $m$ with vertex sequence $(v_0, v_1, \ldots, v_m)$.
According to Eq.~\eqref{eq:IT}, the node influence $I_P(v_i, v_j)$ of any pair of nodes $v_i$ and $v_j$ with a distance of $k$ in path $P$ is
\begin{equation}
\label{eq:IP}
    I_P(v_i, v_j) = \frac{1}{2^{k-1}}, \enspace \text{s.t.}~|i - j| = k
\end{equation}
In addition, there are a total of $(m-k+1)$ pairs of nodes with a distance of $k$ ($1 \le k \le m$) in path $P$.
By using the definition of path information and Eq.~\eqref{eq:IP}, we obtain
\begin{align*}
    I(P)
    &= \sum_{\substack{i,j=0 \\ i \neq j}}^m {I_P(v_i, v_j)} \nonumber \\
    &= 2 \cdot \sum_{k=1}^m {\sum_{i=0}^{m-k} {I_P(v_i, v_{i+k})}} \nonumber \\
    &= \sum_{k=1}^m {(m-k+1) \cdot \frac{2}{2^{k-1}}}
\end{align*}
Thus, finding a path with the maximum path information in tree $T$ is equivalent to solving the following optimization problem:
\begin{align}
    \label{eq:obj}
    \max_{m} \quad
    & I(P) = \sum_{k=1}^m {(m-k+1) \cdot \frac{2}{2^{k-1}}} \\
    \text{s.t.} \quad
    & 1 \le m \le \mathfrak{d} \nonumber \\
    & m \in \mathbb{Z}^+ \nonumber
\end{align}
where $\mathfrak{d}$ is the diameter of tree $T$.

Consider a finite sequence $\left(x_m\right)_{m=1}^\mathfrak{d}$ that corresponds to the objective function Eq.~\eqref{eq:obj}, where
\begin{equation*}
    x_m = \sum_{k=1}^m {(m-k+1) \cdot \frac{2}{2^{k-1}}}
\end{equation*}
Since
\begin{equation*}
    x_{m+1} - x_m = \sum_{k=1}^{m+1} {\frac{2}{2^{k-1}}} > 0, \enspace \forall m \in \mathbb{Z}^+
\end{equation*}
We can conclude that sequence $\left(x_m\right)_{m=1}^\mathfrak{d}$ is strictly increasing.
Accordingly, the optimization problem described in Eq.~\eqref{eq:obj} is equivalent to the following one:
\begin{align*}
    \max_{m} \quad
    & m \\
    \text{s.t.} \quad
    & 1 \le m \le \mathfrak{d} \nonumber \\
    & m \in \mathbb{Z}^+ \nonumber
\end{align*}
Clearly, the optimal solution is $\hat{m} = \mathfrak{d}$, which corresponds to the longest shortest path $\hat{P}$.
In conclusion, the longest shortest path $\hat{P}$ has the maximum path information $I(\hat{P}) = \sum_{k=1}^\mathfrak{d} {(\mathfrak{d}-k+1) \cdot \frac{2}{2^{k-1}}}$.
\end{proof}

\section{Illustration of the HEmoN model}
\label{sec:example}

\begin{figure*}
    \centering
    \includegraphics[width=0.8\textwidth]{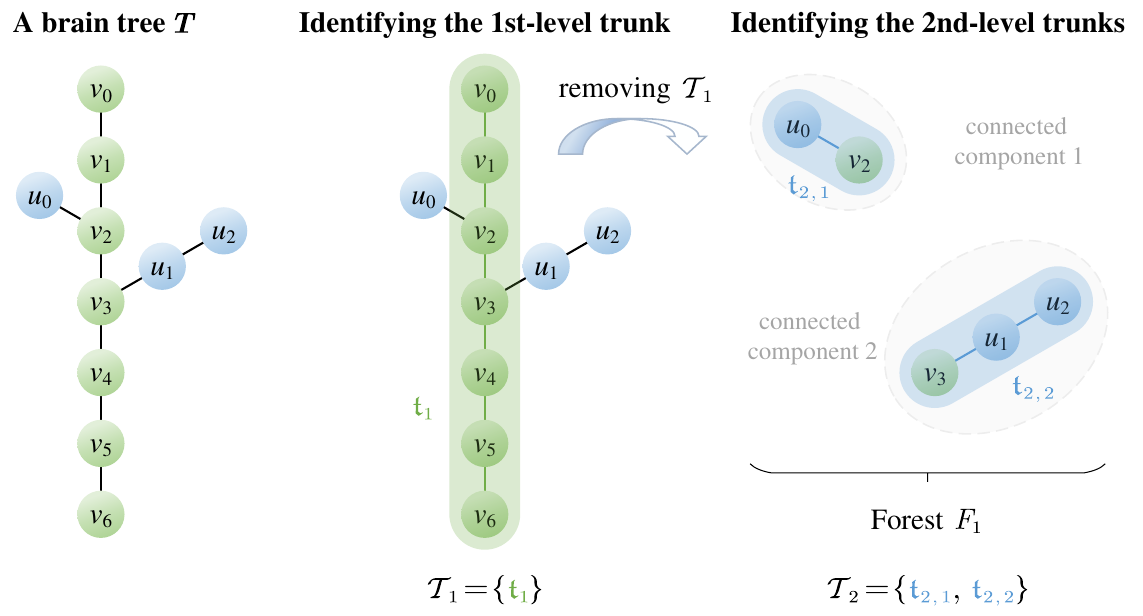}
    \caption{A brain tree (left, adapted from Figure~\ref{fig:pl}) and the process of identifying its hierarchical trunks (middle and right).}
    \label{fig:example}
\end{figure*}

In this section, we illustrate the process of identifying hierarchical trunks and the representation learning process of our proposed HEmoN model through a concrete example, as shown in Figure~\ref{fig:example}.
Let $T = (\mathcal{V}, \mathcal{R})$ denote the brain tree in Figure~\ref{fig:example}, where
\begin{align*}
    \mathcal{V} = \left\{\right.
    & v_0, v_1, \ldots, v_6, u_0, u_1, u_2 \left.\right\}, \\
    \mathcal{R} = \left\{\right.
    & (v_0, v_1), (v_1, v_2), \ldots, (v_5, v_6), \\
    & (u_1, u_2), (v_2, u_0), (v_3, u_1) \left.\right\}.
\end{align*}

According to the methodology described in Section~\ref{sec:area}, we designate the longest shortest path as the main trunk $\mathfrak{t}_1$.
As a result, we obtain the set of 1st-level trunk $\mathcal{T}_1 = \{\mathfrak{t}_1\}$.
Then, we remove $\mathcal{T}_1$ from $T$, and the remaining part becomes a forest $F_1$.
This forest contains two connected components: connected component 1 and connected component 2, with each component being a (sub)tree.
To identify the trunks at the second level, we search for the longest shortest path in each connected component of the forest.
Accordingly, we obtain the 2nd-level trunk $\mathfrak{t}_{2,1}$ in connected component 1, the 2nd-level trunk $\mathfrak{t}_{2,2}$ in connected component 2, and thus the set of 2nd-level trunks $\mathcal{T}_2 = \{\mathfrak{t}_{2,1}, \mathfrak{t}_{2,2}\}$.

Since there are no remaining nodes after removing $\mathcal{T}_2$ from $F_1$, we stop the process of identifying hierarchical trunks and decompose $T$ into two levels of trunks, i.e., $\mathcal{T}_1 = \{\mathfrak{t}_1\}$ and $\mathcal{T}_2 = \{\mathfrak{t}_{2,1}, \mathfrak{t}_{2,2}\}$.

After obtaining all the trunks at each level, we learn the trunk representation $\bm h_T^{(\ell)}$ for each level $\ell$, which also serves as the representation of the emotional area at the $\ell$th level.
According to Eq.~\eqref{eq:trunk}, we have
\begin{align*}
    \bm h_T^{(1)}
    &= \text{LSTM} \left( \bm x_{v_0}^{(\mathfrak{t}_1)}, \bm x_{v_1}^{(\mathfrak{t}_1)}, \cdots, \bm x_{v_6}^{(\mathfrak{t}_1)} \right) \\[0.5em]
    \bm h_T^{(2)}
    &= \text{LSTM} \left( \bm x_{u_0}^{(\mathfrak{t}_{2,1})}, \bm x_{v_2}^{(\mathfrak{t}_{2,1})} \right) + \nonumber \\
    &\phantom{{}={}} \text{LSTM} \left( \bm x_{v_3}^{(\mathfrak{t}_{2,2})}, \bm x_{u_1}^{(\mathfrak{t}_{2,2})}, \bm x_{u_2}^{(\mathfrak{t}_{2,2})} \right)
\end{align*}

Finally, according to Eq.~\eqref{eq:final}, we combine both the trunk representations to create a representation of the brain tree~$T$:
\begin{equation*}
    \bm h_T = \sum_{\ell=1}^2 {\bm W^{(\ell)} \bm h_T^{(\ell)}} = \bm W^{(1)} \bm h_T^{(1)} + \bm W^{(2)} \bm h_T^{(2)}.
\end{equation*}

\bibliographystyle{model1-num-names}

\bibliography{cas-refs}

\end{document}